# Cross-layer Balanced and Reliable Opportunistic Routing Algorithm for Mobile Ad Hoc Networks

Ning Li, Universidad Politenica de Madrid
Jose-Fernan Martinez-Ortega, Universidad Politenica de Madrid
Vicente Hernandez Diaz, Universidad Politenica de Madrid

For improving the efficiency and the reliability of the opportunistic routing algorithm, in this paper, we propose the cross-layer and reliable opportunistic routing algorithm (CBRT) for Mobile Ad Hoc Networks, which introduces the improved efficiency fuzzy logic and humoral regulation inspired topology control into the opportunistic routing algorithm. In CBRT, the inputs of the fuzzy logic system are the relative variance (*rv*) of the metrics rather than the values of the metrics, which reduces the number of fuzzy rules dramatically. Moreover, the number of fuzzy rules does not increase when the number of inputs increases. For reducing the control cost, in CBRT, the node degree in the candidate relays set is a range rather than a constant number. The nodes are divided into different categories based on their node degree in the candidate relays set. The nodes adjust their transmission range based on which categories that they belong to. Additionally, for investigating the effection of the node mobility on routing performance, we propose a link lifetime prediction algorithm which takes both the moving speed and moving direction into account. In CBRT, the source node determines the relaying priorities of the relaying nodes based on their utilities. The relaying node which the utility is large will have high priority to relay the data packet. By these innovations, the network performance in CBRT is much better than that in ExOR; however, the computation complexity is not increased in CBRT.

## 1. INTRODUCTION

The Mobile Ad Hoc Networks (MANETs) are widely used in Internet of Things (IoT), such as in battlefield [1][2], vehicle network [3][4][5], underwater cooperation robot network [6][7][8], etc. In MANETs, one of the important issues is the routing protocol design, which guarantees reliable and efficient data transmission from source node to destination node. There are two main routing strategies for the MANETs: deterministic routing and opportunistic routing [9]. In deterministic routing, the sender sends data packet to one neighbor node which is chosen based on the optimal algorithms; in opportunistic routing, the sender sends the data packet to a set of relaying nodes rather than only one relaying node to improve the packet delivery ratio between sender and receiver [10][11][12][13]. In this paper, we mainly focus on the opportunistic routing.

### 1.1. Motivations

Ning Li, Jose-Fernan Martinez-Ortega, and Vecente Hernandez Diaz are with the Universidad Politenica de Madrid, Madrid, Spain.
The research leading to the presented results has been undertaken with in the SWARMs European project (Smart and Networking Underwater Robots in Cooperation Meshes), under Grant Agreement n. 662107-SWARMs-ECSEL -2014-1, partially supported by the ECSEL JU and the Spanish Ministry of Economy and Competitiveness (Ref: PCIN-2014-022-C02-02).
E-mail: {li.ning, jf.martinez, vicente.hernandez}@upm.es.

In opportunistic routing, the source node chooses the candidate relay nodes and decides the priorities of these nodes based on different performance metrics, such as the distance to the destination node, the expected transmission count (ETX), the propagation delay, the queue length, etc [13]. These performance metrics are relevant and interplay; the more performance metrics are taken into account, the better routing performance is. In traditional opportunistic routing algorithm, since the source node chooses the relaying nodes based on limited number of performance metrics, so it cannot reflect the comprehensive characteristic of the whole network. For instance, when the source node chooses and prioritizes the relaying nodes based on their ETX, the nodes which the ETX is small will be chosen and set with high priorities. However, considering the parameters in MAC layer, link layer, or physic layer, the node which the ETX is small does not mean that the other parameters are also optimal. The effective approach for solving this predicament is to figure out an algorithm which can get the optimal solutions for all the cross-layer metrics at the same time. Unfortunately, this is impossible and has been proved is NP-hard problem. For finding the tradeoff between network parameters, the fuzzy logic has been introduced into the routing algorithm, such as in [14], [15], [16], [17], [18], etc. However, the main disadvantage of fuzzy logic is that with the increasing of the number of inputs, the number of fuzzy rules increases exponential. This means that the fuzzy logic based routing algorithms can only handle limited number of cross-layer parameters at the same time. In general, this number is no larger than 3. This limits the further performance improvement of the fuzzy logic based opportunistic routing algorithms.

In opportunistic routing, the more nodes in candidate relays set [13], the higher packet delivery ratio between the sender and the candidate relays set is; vice versa. However, due to the mobility of the nodes in MANETs, the number of neighbors changes frequently, so the packet delivery ratio between the sender and the candidate relays set changes frequently, too. In case the node number is too small or too large in the candidate relays set, in this paper, we introduce the topology control, which is good at controlling the number of neighbors [19], into the opportunistic routing algorithm. However, to the traditional topology control algorithms, due to the high mobility of nodes in MANETs, if each node maintains constant number of nodes in candidate relays set, the transmission power needs to be adjusted frequently, which consumes a plenty of network resource that could have been used in data packet transmission. So the topology control algorithm should be able to reduce the extra control cost as far as possible. Moreover, since the node mobility has great effection on the routing performance, so the node mobility, including the moving speed and moving direction, should also be taken into account in the routing algorithm design.

**1.2. Main contributions**

Motivated by these, we propose the cross-layer and reliable opportunistic routing algorithm (CBRT) for MANETs in this paper. In CBRT, the inputs of the fuzzy logic system are the relative variance ($rv$) of the metrics rather than the value of the metrics, which can reduce the number of fuzzy rules dramatically. Moreover, the number of fuzzy rules does not increase when the number of inputs increases. For reducing the control cost, the number of nodes in the candidate relays set is a range rather than a constant number in CBRT. The nodes are divided into different regions based on the numbers of nodes in their candidate relays set. The nodes adjust their transmission ranges based on which categories that they belong to. The source node sets the priorities for the nodes in the candidate relays set based on the node utilities. The relaying node which the utility is large has high relaying priority. The main contributions of this paper are summarized as follows:
- We propose a weight based fuzzy logic algorithm (SBFL) which can handle multiple cross-layer parameters at the same time without increasing the computation complexity; to the best of our knowledge, this is the first algorithm which can take so many cross-layer parameters into account without increasing the computation complexity seriously;

- We propose a link lifetime prediction algorithm which takes not only the moving speed but also the moving direction into consideration to predict the link lifetime;
- We propose an packet delivery ratio based opportunistic topology control algorithm for maintaining stable node number in the candidate relays set; in this algorithm, the node number in the candidate relays set is stable and the transmission power adjustment ratio is low;
- Based on the topology control algorithm and SBFL algorithm, we propose the cross-layer and reliable opportunistic routing algorithm, which utilizes both the SBFL and topology control to improve the routing performance.

The remaining of this paper organized as follows: in Section II, we introduce the related works; Section III introduces the principle of SBFL; the link lifetime prediction algorithm and the packet delivery ratio based topology control algorithm are presented in Section IV, the CBRT algorithm is also introduced in this section; Section V evaluates the performance of the CBRT algorithm and compares the performance of CBRT algorithm with that of the ExOR algorithm; Section VI concludes this paper.

## 2. RLATED WORKS

Since the opportunistic routing has been proposed by Sanjit Biswas and Robert Morris [20], there are many opportunistic routing algorithms have been developed in recent years. In [21], the authors propose a relaying node selection algorithm for 1-D wireless sensor network, named energy saving via opportunistic routing (ENS_OR). The ENS_OR selects the candidate relays set and prioritizes the nodes in it based on their virtual optimal transmission distance and residual energy level. The nodes which are closer to the energy equivalent nodes and have more residual energy than the source node will be selected as the relaying nodes. In [22], the authors propose two accurate energy consumption based objective functions, which exploit the knowledge of both the frame error ratio within the physical layer, the maximum number of retransmissions in the medium access control (MAC), and the number of relays in the network layer. Based on the objective functions, the routing algorithm is designed for opportunistic routing, which employs the objective functions to prioritize the nodes in the candidate relays set. The authors in [23] exploit the geographic opportunistic routing (GOR) for Quality of Services (QoS) provisioning with both the end to end reliability and delay constrains in wireless sensor networks (WSNs); moreover, the authors also define the problem of efficient GOR for multiconstrained QoS provisioning in WSNs. Based on these conclusions, the authors propose the Efficient QoS-aware GOR protocol for the WSNs, in which the energy efficiency, latency, and the time complexity are taken into account to select the relaying nodes. The authors in [24] propose a distributed optimal community-aware opportunistic routing algorithm, in which the home-aware community model is proposed. In this algorithm, first, the routing between lots of nodes are turned to the routing between a few community homes; then the algorithm maintains an optimal candidate relays set for each home. Each home only forwards its message to the node in its optimal candidate relays set. More opportunistic routings can be found in [25], [26] and [13]; especially in [13], the author reviews the opportunistic routing algorithms in recent years and identifies and discusses the future research directions related to the opportunistic routing design, optimization, and deployment.

The fuzzy logic becomes popular recently, many fuzzy logic based routing protocol have been proposed. In [27], the authors propose an energy-effective cross-layer routing protocol for WSNs based on fuzzy logic. In this protocol, for minimizing the energy consumption and maximizing the network lifetime, the algorithm takes the remaining battery reserve capacity, the link quality, and the transmission power of the neighbor nodes into account to select the next hop relaying nodes. In [28], for reducing the average end-to-end delay of the MANETs, the authors propose a fuzzy logic based adaptive cross-layer routing protocol for the delay-sensitive applications. In this algorithm, each node can switch between reactive routing mode and proactive routing mode based on the current node status separately. The algorithm uses the fuzzy logic controller to decide the routing mode of each node. In [29], the authors propose a routing algorithm for WSNs which can extend the network lifetime and balance the energy consumption by combining the fuzzy approach and the A-star algorithm together. The remaining battery power, the number of hops to the destination node, and the traffic loads are taken into consideration to determine the optimal routing path from the source node to the destination node. Similar in [30], to prolong the network lifetime, a fuzzy logic based energy-optimization routing protocol is proposed, in which the social welfare function is used to predict inequality of residual energy of neighbor nodes. The algorithm computes the degree of energy balance based on the energy inequality. The fuzzy logic system uses the degree of node closeness to the shortest path, the degree of node closeness to sink, and the degree of energy balance to achieve the routing decision. Additionally, in [31], the node density, the delay, and the number of dead nodes are used as inputs of the fuzzy logic system to select the next hop relaying node for achieving the balanced energy consumption of all the nodes with minimum delay. More related works about the fuzzy logic based routing algorithms can also be found in [32], [33], [34], and [35].

Recently, more and more researchers consider that integrate the topology control and routing algorithm in wireless sensor network or ad hoc network is an effective way to improve the network performance. In [36], the authors proposed three alternative mathematical models for integrating topology control and routing decisions so as to prolong the lifetime of sensors. For reducing the effection of primary user (PU) activities and node mobility to the stability of links in mobile cognitive networks, in [37], the authors proposed a primary user activity prediction based joint topology control and stable routing protocol (PP-JTCSR), which can quantitatively capture channel utilization patterns of PUs. By topology control in PP-JTCSR, the most stable and shortest path can be found. In [38], the authors investigate the impact of two topology control methods for resolving the problem of void/isolated nodes appeared in geographic routing protocols, which can reduce the number of void/isolated nodes significantly. In [39], the authors have shown that joint topology control and routing assignment as an optimization problem is a NP-hard problem. For solving this problem, in this paper, the author proposed TORA (joint topology control and routing assignment) which seeks to jointly optimize topology and routing for DMesh (directional antennas in wireless mesh networks). There are also many researches that integrate the routing design with the topology control, which can be found in [40], [41], [42], and [43].

## 3. PRINCIPLE OF THE SCATTER BASED FUZZY LOGIC ALGORITHM

In this section, the principle of the weight based fuzzy logic algorithm (SBFL) will be introduced in detail.

### 3.1. Problem statement

As shown in Fig. 1, the fuzzy logic system is composed by three modules: fuzzification, fuzzy logic inference, and defuzzification.

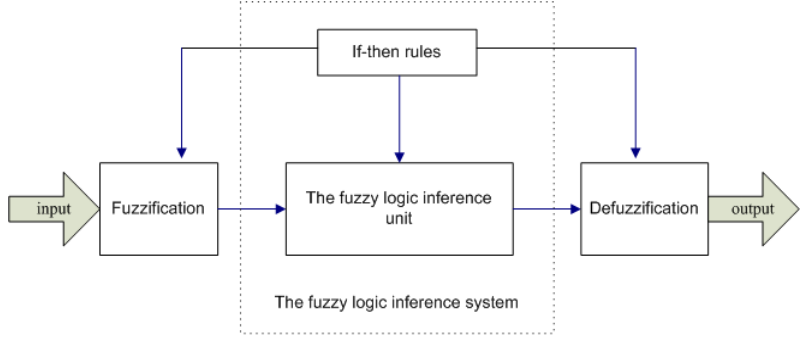

Fig. 1. The process of fuzzy logic system

In fuzzification module, the input universes are mapped to fuzzy set based on the linguistic variables and membership functions. Assuming that the input universe is $U$ and the fuzzy set of universe $U$ is $A$; $\mu_A$ is the membership function which maps universe $U$ to fuzzy set $A$. The membership function represents the membership between universe $U$ and fuzzy set $A$: $\mu_A: U \to A \in [0,1]$. The fuzzy set $A$ can be expressed as: $A = \{(u_i, \mu_A(u_i))|u_i \in U\}$, where $u_i$ is the element of universe $U$. The general definition of membership function is:

$$\begin{cases} \mu_A(u_i) = 1: \mu_i \text{ belongs to A Complete}; \\ \mu_A(u_i) = 0: \mu_i \text{ does not belong to A Complete}; \\ 0 < \mu_A(u_i) < 1: \mu_i \text{ belongs to A partly}; \end{cases} \quad (1)$$

The fuzzy set $A$ is the input of the fuzzy inference module which will be used to calculate the output fuzzy set $B$. The core part of the fuzzy inference system is the if-then rules (fuzzy rules). The if-then rules determine the relationship between the inputs (fuzzy set $A$) and outputs (fuzzy set $B$). The form of the fuzzy rule is generally expressed as:

$$\text{If } A \text{ (input) then } B \text{ (output)} \quad (2)$$

where $A$ and $B$ are the fuzzy sets of universe $X$ (input) and $Y$ (output), respectively. The membership functions are $\mu_A(x_i)$ and $\mu_B(x_i)$, where $x_i \in X$, $y_i \in Y$. Thus according to the fuzzy mathematic theory (Mamdani fuzzy inference algorithm), the membership function between fuzzy set $A$ and fuzzy set $B$ can be decided by:

$$\mu_{A \to B}(x, y) = \mu_R(x, y) \triangleq \mu_A(x) \wedge \mu_B(y) \quad (3)$$

where $R$ represents the fuzzy relationship between universe $X$ and universe $Y$, i.e. $R = A \to B$; "$\wedge$" is the conjunction operator in discrete mathematics. By (3), the fuzzy result can be got.

However, for taking more cross-layer performance metrics into account, the fuzzy rules will be much more complex than that shown in (2), which is a multi-inputs-single-output fuzzy system. In this scenario, the fuzzy rules will be:

$$\text{If } A_1, A_2, A_3, \ldots, A_n \text{ then } B \quad (4)$$

where $A_1, A_2, A_3, \ldots, A_n$ (inputs) and $B$ (output) are the fuzzy sets of universes $U_1, U_2, U_3, \ldots, U_n$ (inputs) and $U$ (output), respectively. The membership functions are $\mu_A(u_1), \mu_A(u_2), \mu_A(u_3), \ldots, \mu_A(u_n), \mu_B(u)$, where $u_1 \in U_1, u_2 \in U_2, u_3 \in U_3, \ldots, u_n \in U_n$, and

$u \in U$. So the membership function can be expressed as:

$$\mu_{\{A_1,A_2,A_3,\ldots,A_n\}\to B}(u_1, u_2, u_3, \ldots, u_n, u) = \mu_R(u_1, u_2, u_3, \ldots, u_n, u) \\ \triangleq \mu_{A_1}(u_1) \wedge \mu_{A_2}(u_2) \cdots \mu_{A_n}(u_n) \wedge \mu_B(u) \quad (5)$$

where $R = \{A_1, A_2, A_3, \ldots, A_n\} \to B$ is the fuzzy relationship between universe $U_1$, $U_2$, $U_3$, …, $U_n$ and $U$.

As shown in (5), in multi-inputs-single-output fuzzy system, the number of the permutation and combination between the fuzzy sets are huge. The (5) illustrates that with the increasing of the fuzzy set, the calculation complexity of the fuzzy logic system increases seriously. Moreover, if we take the number of linguistic variables into account, the calculation is too complexity to be accepted. For instance, if the number of linguistic variables is *n* and the number of fuzzy sets is *m*, then according to (5), the total number of the fuzzy rules is $n^m$. This means that with the increasing of either the fuzzy set or the linguistic variables, the number of the fuzzy rules will increase exponential. The large number of fuzzy rules needs large memory space of nodes, which is always impractical in MANETs. For solving this issue, we propose an algorithm which can isolate the effection of the number of inputs from the fuzzy inference system, i.e. the number of fuzzy rules does not increase when the number of inputs increases. The new algorithm is named weight based balanced fuzzy logic algorithm, shorted as SBFL.

### 3.2. Principle of the SBFL algorithm

Considering the fact that during the relaying node selection and prioritization, the performance metric (such as the residual energy, the ETX, the distance to the destination node, etc) which the variation rate between different nodes is larger has greater effection on the routing performance than that of the metric which the variation rate is small. For instance, as the performance metrics shown in Table 1, in which the variation rate of Metric_1 is smaller than that of Metric_2. During the relaying node selection and prioritization, from the point of view of Metric_1, which node is chosen as the next hop relaying node has small effection on the routing performance, since the values of Metric_1 in these three nodes are similar; however, considering the values of Metric_2, which node is chosen has greater effection on the routing performance than that of Metric_1. For evaluating the variation rate of the metrics, one of the available metric is the variance. However, as shown in Table 1, the variance is affected seriously by the value of the metric, which can be found in Table 1. In Table 1, the variance of Metric_1 is larger than that of Metric_2; however, taking the values of the metrics into account, the variation rate of Metric_1 is smaller than that of Metric_2 in fact. So a new metric should be developed for reflecting the accurate variation rate of the performance metrics.

TALBE 1. METRICS AND PRIORITIES

|  | node1 | node2 | node3 | V | rv | weight |
|---|---|---|---|---|---|---|
| Metric_1 | 1001 | 1002 | 1003 | 0.667 | 6.64×10$^{-6}$ | 0.0523 |
| Order of Metric_1 | 1 | 2 | 3 |  |  |  |
| Metric_2 | 0.5 | 0.8 | 0.1 | 0.0822 | 0.377 | 0.753 |
| Order of Metric_2 | 2 | 3 | 1 |  |  |  |
| Metric_3 | 1000 | 2000 | 3000 | 666667 | 0.167 | 0.333 |
| Order of Metric_3 | 1 | 2 | 3 |  |  |  |
| Node utility | 1.891 | 2.697 | 1.909 |  |  |  |

In SBFL algorithm, assuming that there are *n* input universes, which is $U = [U_1, U_2, \ldots U_n]$; for each universe, there are m elements, i.e. $U_i = [U_i^1, U_i^2, \ldots, U_i^m]^T$; therefore, $U$ is a $m \times n$ matrix:

$$\begin{bmatrix} U_1^1, U_2^1, \dots, U_n^1 \\ U_1^2, U_2^2, \dots, U_n^2 \\ \vdots \\ U_1^m, U_2^m, \dots, U_n^m \end{bmatrix}.$$

For developing the more accurate metric, we propose the relative variance (*rv*) which takes the average value of the metrics into account to evaluate the variation rate of the metric, shown as:

$$D_i = \frac{1}{m}\sum_{j=1}^{m}\left(\frac{U_i^j - \bar{U}_i}{\bar{U}_i}\right)^2, i \in [1,n] \tag{6}$$

where $U_i^j$ means the *jth* element of universe $U_i$; $\bar{U}_i$ is the mean value of the parameters in $U_i$ and can be calculated as:

$$\bar{U}_i = \frac{1}{m}\sum_{j=1}^{m} U_i^j, i \in [1,n] \tag{7}$$

In (6), the variation rate will be not affected by the value of the metric, which can be found in Table 1. In Table 1, even the variance of Metric_1 is larger than that of Metric_2, the rv of Metric_1 is much smaller than that of the Metric_2, which can reflect the variation rate of the metric more accurate than variance.

In SBFL, the input universes $U_1$, $U_2$, …, $U_n$ are replaced by the relative variance $D$ as the input of the fuzzy inference system. The fuzzy set $A_D$ can be expressed as:

$$\begin{aligned} A_D &= \{(d, d_A(d_i)) | d_i \in D\} \\ &= [d_A(d_1), d_A(d_2), \dots, d_A(d_n)] \end{aligned} \tag{8}$$

where $d_A(d_i)$ is the membership function between relative variance $D$ and fuzzy set $A_D$. Since $D$ is a $1 \times n$ matrix, so the fuzzy rules shown in (5) can be rewritten as: If $A_D$ then $B$. Therefore, the membership function between $A_D$ and $B$ will be:

$$\mu_{D \to B}(z, y) = \mu_R(z, y) \triangleq \mu_D(z) \wedge \mu_B(y) \tag{9}$$

As shown in (9), the membership function in (9) is much simpler than that in (5). By using the concept of relative variance, the number of fuzzy set has been reduced from n to 1, so the calculation complexity is reduced. Moreover, the number of fuzzy rules in SBFL does not increase when the number of fuzzy set and the linguistic variables increase. Additionally, as shown in Fig. 2, the complexity caused by the increasing of the number of inputs has been isolated from the fuzzy inference system successfully. This means no matter how many different universes are inputted into SBFL, the number of input is only one. Therefore, in MANETs, the SBFL algorithm can take as many cross-layer metrics into account as possible to figure out the most efficient solution without increasing the computation complexity.

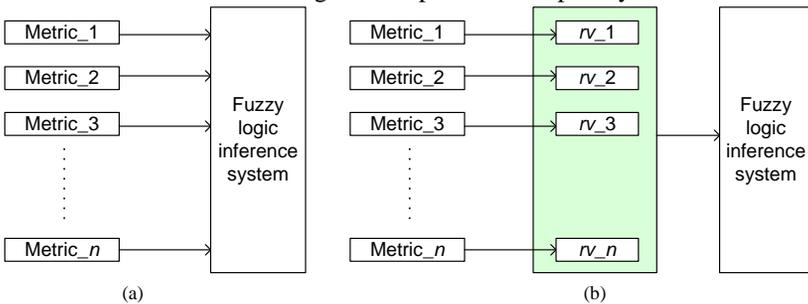

Fig. 2. (a) The principle of traditional fuzzy logic system; (b) The principle of SBFL algorithm.

The outputs of SBFL are the weights of the cross-layer metrics:

$$\omega = [\omega_1, \omega_2, \omega_3, \ldots, \omega_m] \quad (10)$$

In SBFL, since the cross-layer metrics whose *rv* are large have great effection on the relay node selection; conversely, the metrics whose *rv* are small have small effection on the relay node selection; therefore, the weights of the metrics whose *rv* are large should large; otherwise, the weights are small. Moreover, as the number of the linguistic variables in SBFL can be set as large as possible, so according to [14], the number of the linguistic variables is set to 7 in this paper. The fuzzy linguistic variables and the fuzzy rules are shown in Table 2.

TABLE 2. FUZZY LOGIC RULES

| **IF** Input (*D*) | **THEN** Output ($\omega$) |
|---|---|
| very small | very small |
| medium small | medium small |
| small | small |
| medium | medium |
| large | large |
| medium large | medium large |
| very large | very large |

When the weight of each cross-layer metric has been gotten, which are shown in Table 1, the $\omega$ and *U* will be used to calculate the utilities of nodes in the candidate relays set:

$$\psi = \omega \times U = [\psi_1, \psi_2, \psi_3, \ldots, \psi_n] = \left[\sum_{j=1}^{m} \omega_j U_i^j, \sum_{j=1}^{m} \omega_j U_2^j, \ldots, \sum_{j=1}^{m} \omega_j U_n^j\right] \quad (11)$$

However, the important issue should be paid attention in SBFL is that the order-of-magnitudes of the metrics may different, which have great effection on the algorithm. If we use the values of these metrics directly in SBFL algorithm, there will have mistakes. For instance, considering the metrics which are shown in Table 1; in Table 1, the Metric_3 is much bigger than Metric_2 and Metric_1. If the SBFL algorithm is applied directly, the utilities of node1, node2, and node3 are 385.73, 719, and 1051.54, respectively. As a result, the priorities of these nodes are: *node3→node2→node1*. However, since the value of Metric_3 is much larger than the other two metrics, so the node utilities in Table 1 are decided mainly by the Metric_3. The Metric_3 will cover up the effection of Metric_1 and Metric_2 on relay node selection. The reason is that the metrics in Table 1 are not in the same order-of-magnitudes; therefore, the parameters whose order-of-magnitudes are larger than others will have great effection on the SBFL algorithm.

In this paper, for solving this issue, we introduce the sequence of the parameter in metric into the calculation of the node utility. The sequence of the parameter can reflect the relationship of these parameters. For instance, as the performance metrics shown in Table 1, the sequence of the parameter is shown in Table 1. The priority of Metric_i is denoted as $R_{Metric\_i}$. So the (11) can be rewritten as:

$$\psi = \omega \times U = [\psi_1, \psi_2, \psi_3, \ldots, \psi_n] = \left[\sum_{j=1}^{m} \omega_j R_{Metric\_j}^1, \sum_{j=1}^{m} \omega_j R_{Metric\_j}^2, \ldots, \sum_{j=1}^{m} \omega_j R_{Metric\_j}^n\right] \quad (12)$$

where $R_{Metric\_j}^1$ is the sequence of the parameter of Metric_j in node 1; for instance, in Table 1,

$R^1_{Metric\_1} = 1$, $R^1_{Metric\_2} = 2$, and $R^1_{Metric\_3} = 1$.

According to (12), the utilities of these three nodes are shown in Table 1. By using (12), the node priorities will be: *node2→node3→node1*, which is more reasonable than that of the node utilities that calculated by (11). As shown in Table 1, the orders of the node utilities are decided by both of these three performance metrics.

A simple example can be found as follows. In this example, we evaluate the effectiveness of SBFL algorithm. The parameters are shown in Table 3. There are five nodes. In each node, there are three metrics.

TABLE 3. AN EXAMPLE

|  | node1 | node2 | node3 | node4 | node5 | rv | weight |
|---|---|---|---|---|---|---|---|
| metric_1 | 45.1 | 84 | 22.9 | 91.3 | 15.2 | 0.361 | 0.849 |
|  | 3 | 4 | 2 | 5 | 1 |  |  |
| metric_2 | 0.602 | 0.263 | 0.654 | 0.689 | 0.784 | 0.0914 | 0.233 |
|  | 2 | 1 | 3 | 4 | 5 |  |  |
| metric_3 | 826 | 538 | 996 | 78 | 443 | 0.305 | 0.76 |
|  | 4 | 3 | 5 | 1 | 2 |  |  |
| Utility1[1] | 6.053 | 5.909 | 6.197 | 5.931 | 3.534 |  |  |
| Utility2[2] | 0.501 | 0.467 | 0.527 | 0.5 | 0.44 |  |  |

[1] the node utility calculated by (12);
[2] the node utility calculated by the traditional fully logic algorithm, which the number of linguistic variables is 3 (small, medium, large);

Table 3 shows the performance of the traditional fuzzy logic algorithm and SBFL algorithm on prioritizing the nodes, respectively. Table 3 demonstrates that based on SBFL algorithm, the utilities of node1, node2, and node4 are similar, which are higher than node5; the utility of node3 is the largest in these five nodes. Therefore, the priorities of all the nodes are: node3→node1→node4→ node2→node5. The priorities of the nodes that decided by the traditional fuzzy logic algorithm are also presented in Table 3: node3→node1→ node4→node2 →node5, which is the same as the node utilities that calculated by (12). This demonstrates that the approach shown in (12) is effective on determining the relaying priorities of the relaying nodes.

Another disadvantage of the traditional fuzzy logic algorithm is the calculation complexity. In this simulation, to the traditional fuzzy logic algorithm, the number of the parameters is 3 in each node and the number of the linguistic variable is 3, so the number of fuzzy rules is 27; however, in the SBFL algorithm, this number is only 7, which is equal to the number of the linguistic variables. This can be found in Fig. 3. In Fig. 3(a), the number of the linguistic variable is 3 in traditional fuzzy logic system and 7 in SBFL algorithm. With the increasing of the fuzzy sets, the number of the fuzzy rules in traditional fuzzy logic algorithm increases exponential; however, this number will keep constant in SBFL algorithm. The similar result can be found in Fig. 3(b). In Fig. 3(b), the number of the fuzzy set is fixed and equal to 3. With the increasing of the linguistic variables, the fuzzy rules in traditional fuzzy logic system increases exponential like that in Fig. 3(a); however, in SBFL algorithm, the increasing is slightly and equal to the number of the linguistic variables. Fig. 3 demonstrates that the SBFL algorithm have great advantages on dealing with the multi-inputs issues.

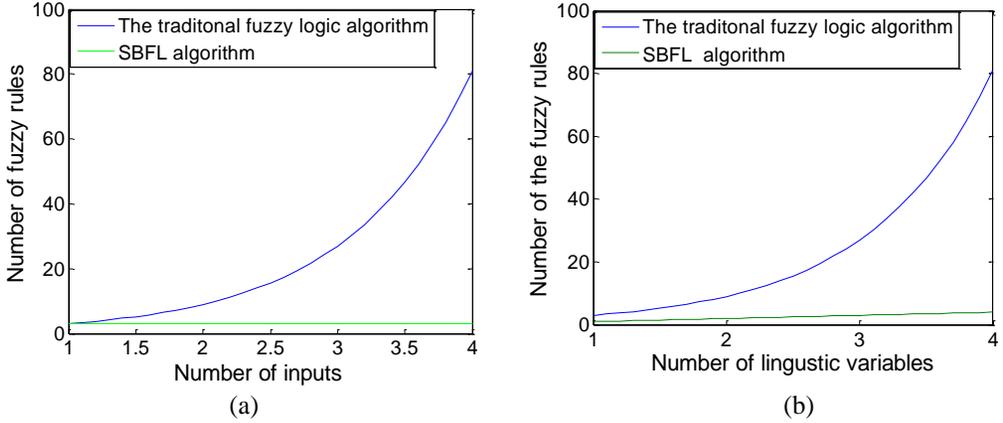

Fig. 3. The number of rules: (a) the number of linguistic variables is fixed; (b) the number of fuzzy set is fixed.

## 4. CROSS-LYAER BALANCED AND RELIABLE OPPORTUNISTIC ROUTING ALGRITHM

Based on the SBFL algorithm that proposed in Section III, we propose the cross-layer and reliable opportunistic routing algorithm (CBRT) for MANETs in this section. Considering the mobility of the nodes in MANETs, in this section, the topology control algorithm and the link lifetime prediction algorithm which takes both the moving speed and the moving direction into account are developed.

### 4.1. Link lifetime prediction algorithm

In MANETs, the routing performance is affected by node mobility seriously. Due to the node mobility, the network topology changes frequently, so the router easily to be broken and cannot last a long time [44]. The link lifetime is affected by the node mobility seriously. The node mobility includes the moving speed and moving direction. Both the moving speed and moving direction can affect the link lifetime. However, in the previous researches, the effection of the moving direction on the link lifetime has not been fully investigated. In this section, this issue will be investigated in detail.

The link lifetime prediction is difficult in MANETs because the nodes can move freely [44]. During the calculation, we assume that the link connections are unstable and can only last a short period of time. Moreover, in CBRT, each node knows other nodes' location and only the neighbors whose distances to the destination node are smaller than that of the source node will be considered as the candidate relaying nodes. As shown in Fig. 4, when the nodes locate in the red area (which is defined as the survival area, the survival area means that in this area, the link connection can be guaranteed), the link lifetimes are larger than 0. The value of the link lifetime relates to the source node $s$, the relaying node r, and the destination node d. Due to the mobility of the source node $s$ and the destination node $d$, the survival area changes continuously. In addition, considering the mobility of the relaying node $r$, the communication link will be broken when the node $r$ moves out of the survival area. Therefore, the relative velocity of node $r$ relative to node $s$ and node $d$ need to be calculated. The calculation of the relative velocity is shown in Fig. 5.

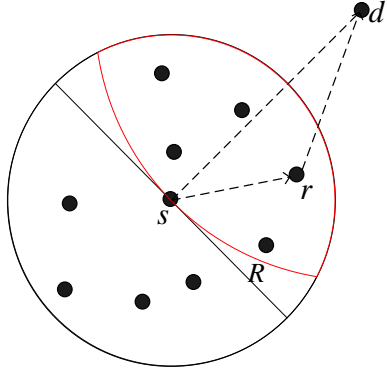

Fig. 4. Geographic based relay node selection

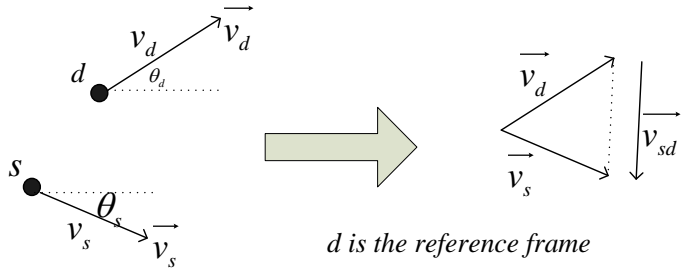

*d is the reference frame*

Fig. 5. The principle of the velocity vector operation

In Fig. 5, the moving velocities of node *s*, node *r*, and node *d* are $\vec{v_s}$ (the moving speed is $v_s$ and the moving direction is $\theta_s$), $\vec{v_r}$ (the moving speed is $v_r$ and the moving direction is $\theta_r$), and $\vec{v_d}$ (the moving speed is $v_d$ and the moving direction is $\theta_d$), respectively. If the destination node d is chosen as the reference frame, then the relative velocity of source node s relative to node *d* is given by: $\vec{v_{sd}} = \vec{v_s} - \vec{v_d}$. According to the vector synthesis theory, the relative moving speed and moving direction of $\vec{v_{sd}}$ can be given by: $v_{sd} = \sqrt{v_{xsd}^2 + v_{ysd}^2}$, $\theta_{sd} = \arctan(v_{xsd}/v_{ysd})$, $v_{xsd} = v_s \cos\theta_s + v_d \cos\theta_d$, and $v_{ysd} = v_s \sin\theta_s + v_d \sin\theta_d$; where $v_{xsd}$ is the moving speed of $v_{sd}$ in *x*-axis and $v_{ysd}$ is the moving speed of $v_{sd}$ in *y*-axis. Therefore, the relative velocity of relay node *r* relatives to $\vec{v_{sd}}$ can be calculated as: $\vec{v_{sdr}} = \vec{v_r} - \vec{v_{sd}}$. Similarly with that shown above, the relative moving speed and moving direction of $\vec{v_{sdr}}$ can be got from: $v_{sdr} = \sqrt{v_{xsdr}^2 + v_{ysdr}^2}$, $\theta_{sdr} = \arctan(v_{xsdr}/v_{ysdr})$, $v_{xsdr} = v_r \cos\theta_r + v_{ds} \cos\theta_{sd}$, and $v_{ysdr} = v_r \sin\theta_r + v_{sd} \sin\theta_{sd}$, where $v_{xsdr}$ is the speed of $v_{sdr}$ in *x*-axis and $v_{ysdr}$ is the speed of $v_{sdr}$ in *y*-axis.

As shown in Fig. 6, node r moves with velocity $\vec{v_{sdr}}$. Assuming that the node *r* moves out of the survival area at time *t*, according to Fig. 6, there are two different scenarios about this issue: (1) the relaying node moves toward to the destination node; (2) the relaying node moves far away from the destination node. For solving this problem, we need calculate the relative velocity angle of node *r*.

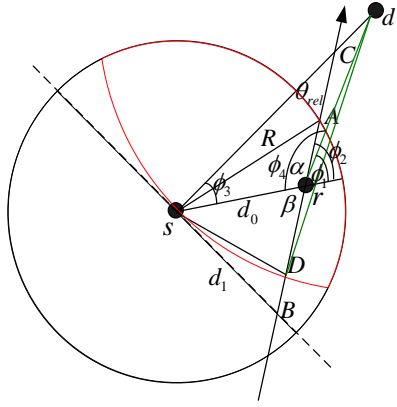

Fig. 6. The principle of the residual link lifetime calculation

As shown in Fig. 6, assuming the coordinate of node *s* is $(x_s, y_s)$ and the coordinate of node *d* is $(x_d, y_d)$, then the angle of $\overline{sd}$ relative to the *x*-axis can be got from: $\theta_{\overline{sdx}} = \arctan\left(\frac{y_s - y_d}{x_s - x_d}\right)$. According to $\theta_{sdr} = \arctan(v_{xsdr}/v_{ysdr})$ and the triangle geometry theory, the velocity angle of node *r* relates to $\overline{sd}$ is: $\theta_{rel} = \theta_{sdr} - \theta_{\overline{sdx}}$. Therefore, if $\theta_{rel} \in \left[0, \frac{\pi}{2}\right] \cup \left[\frac{3\pi}{2}, \pi\right]$, the node *r* moves close to the destination node in *x*-axis or *y*-axis; otherwise, when $\theta_{rel} \in \left[\frac{\pi}{2}, \frac{3\pi}{2}\right]$, the node *r* moves far away from the destination node in both *x*-axis and *y*-axis. So the link lifetime prediction should be divided into two different parts.

1. When $\theta_{rel} \in \left[0, \frac{\pi}{2}\right] \cup \left[\frac{3\pi}{2}, \pi\right]$

In this scenario, the relay node *r* moves toward to the destination node *d* in *x*-axis or *y*-axis. According to Fig. 5 and the law of cosines, in triangle *srd* we can get the function as follows:

$$d_{dr}^2 = d_0^2 + d_{ds}^2 - 2d_0 d_{ds} \cos\phi_3 \tag{13}$$

where $d_{dr}$ is the distance between relay node *r* and destination node *d*; $d_0$ is the distance between source node *s* and relay node *r* at time $t_0$; $d_{ds}$ is the distance between source node *s* and destination node *d*. Since the values of $d_{dr}$, $d_0$, and $d_{ds}$ can be calculated based on the coordinates of these nodes, so the value of $\phi_3$ can be got from (13). Therefore, $\phi_2$ can be calculated as: $\phi_2 = \pi - \phi_3$.

Additionally, in triangle srC, $\theta_{rel}$ and $\phi_3$ are all known, so the angle $\alpha$ can be calculated: $= \pi - \phi_3 - \theta_{rel}$. In triangle *srA*, according to the law of cosines, we can get:

$$R^2 = d_0^2 + [v_{sdr}(t - t_0)]^2 - 2d_0 v_{sdr}(t - t_0)\cos\alpha \tag{14}$$

where *R* is the transmission range of source node *s*. Therefore, the *t* can be calculated by (14).

Thus in this scenario, the residual link lifetime is:

$$T_r = t - t_0 \tag{15}$$

2. When $\theta_{rel} \in \left[\frac{\pi}{2}, \frac{3\pi}{2}\right]$

When $\theta_{rel} \in \left[\frac{\pi}{2}, \frac{3\pi}{2}\right]$, the link lifetime predication is similar to that when $\theta_{rel} \in \left[0, \frac{\pi}{2}\right] \cup \left[\frac{3\pi}{2}, \pi\right]$.

The distance between the destination node d and relay node $r$ at time $t_0$ is $d_{dr}$ and the length of line $dD$ is equal to $d_{ds}$. Moreover, in triangle $drD$, the $\angle drD$ can be calculated as:

$$\angle drD = \phi_5 = \pi - (\phi_2 - \phi_1) \tag{16}$$

Additionally, in triangle *srd*, according to the law of cosines we can get:

$$d_{ds}^2 = d_0^2 + d_{dr}^2 - 2d_0 d_{dr} \cos\phi_1 \tag{17}$$
$$\phi_1 = \pi - \phi_4 \tag{18}$$

In (16), the $\phi_2$ can be calculated by $\phi_2 = \pi - \phi_3$ and $\phi_1$ can be calculated by (17) and (18), so in triangle *drD*, we have:

$$d_{ds}^2 = d_{dr}^2 + [v_{sdr}(t - t_0)]^2 - 2d_{dr} v_{sdr}(t - t_0)\cos\phi_5 \tag{19}$$

In (19), the *t* can be calculated. So the residual link lifetime in this scenario can be calculated by (15).

**4.2. Packet delivery ratio based opportunistic topology control algorithm**

In this section, we propose the packet delivery ratio based opportunistic topology control algorithm (OTC). The OTC is inspired by the humoral regulation, in which the feasible solution is a range rather than a single value. In OTC, the transmission range is adjusted dynamically based on the packet delivery ratio for maintaining the stability and reliability packet delivery ratio; moreover, the OTC reduces the topology control cost successfully by divided the nodes into different categories. Before proposing the topology control algorithm, we first define the necessary definitions which will be used in this section.

*Definition 4.1*: The relay node degree (RND) is defined as the number of neighbors whose distances to the destination node are smaller than that of the source node, i.e. the number of nodes in the survival area (shown in Fig. 4).

*Definition 4.2*: The transmission range adjustment ratio is defined as the ratio of the number of nodes which adjust their transmission range to the number of nodes of the whole network, which can be expressed as: $C_{OTC} = \frac{\text{nodes adjust the transmission power}}{\text{total nodes in the network}}$.

In opportunistic routing algorithm, the packet delivery ratio between the sender and the candidate relays set can be calculated as [20]:

$$PTP = 1 - (1 - p_i)^n \tag{20}$$

where $p_i$ is the packet delivery ratio between sender and one neighbor node; it can be determined by periodically beacon packet exchanging between sender and its neighbors [45][46]; n is the RND of the candidate relays set. According to (20), to different RND, the packet delivery ratio varies greatly, especially in MANETs. For getting stable packet delivery ratio, the RND should keep constant in the candidate relays set. However, due to the high mobility of MANETs, the network topology changes frequently. So if each node maintains constant RND in candidate relays set, the transmission range needs to be adjusted frequently and the transmission range adjustment ratio is high. The high adjustment ratio makes the communication channel more and more congested. Therefore, for reducing the adjustment ratio, the nodes are divided into different categories in OTC according to their packet delivery ratios. Whether the transmission ranges need to be adjusted or not will be decided by which categories that the nodes belong to. The definitions of different categories are shown in Definition 3.

*Definition 4.3.* When the packet delivery ratio of node is in the region $(P_1, P_2)$; then the node is healthy; otherwise, the node is unhealthy.

The definition is shown in Fig. 7 and expressed in (21).

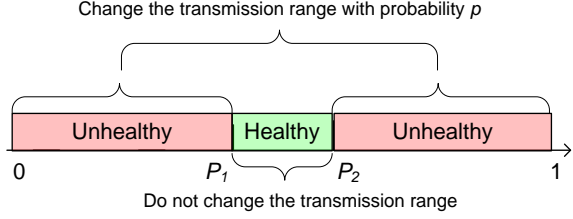

Fig. 7. The definition of the different categories

$$\begin{cases} R_H \in (PTP|PTP \in (P_1, P_2)) \\ R_{UH} \in (PTP|PTP \in (0, P_1) \cup (P_2, 1)) \end{cases} \quad (21)$$

where $P_1$ and $P_2$ are the boundary values of different categories; $R_H$ means the healthy region; $R_{UH}$ means the unhealthy region. To the specific packet delivery ratio, the needed RND of the candidate relays set can be calculated by (20). So the (21) can also be expressed by RND, which is:

$$\begin{cases} R_H \in (n|n \in (n_1, n_2)) \\ R_{UH} \in (n|n \in (0, n_1) \cup (n_2, N)) \end{cases} \quad (22)$$

where $n$ is RND, $N$ is the node number in the network.

The source node calculates the RND in the candidate relays set and adjusts the transmission range according to the value of RND. For reducing the transmission range adjustment ratio, in OTC, the transmission range adjustment probabilities are different in different regions. In health region, this probability is 0; however, in unhealthy region, this probability varies from 0 to 1 based on the value of RND, which is:

$$\begin{cases} p_h = 0, n \in (n_1, n_2) \\ p_u = 1, n \in (0, n_1) \cup (n_2, N) \end{cases} \quad (23)$$

where $p_h$ is the transmission range adjustment probability of healthy region; $p_u$ is the probability of unhealthy region.

In unhealthy region, the adjustment probability varies and is decided by the deviation of the RND in unhealthy region relative to that in healthy region. When the RND is far from the healthy region, the adjustment probability should be high to guarantee the network connectivity and reliability, vice versa. So the adjustment probability can be calculated as:

$$p_{si} = \begin{cases} \frac{n_1 - n_i}{n_1}, n_i \in (n_1, n_2) \\ \frac{n_i - n_2}{N - n_2}, n_i \in (n_2, N) \end{cases} \quad (24)$$

where $n_i$ is the RND of node $i$; $p_{si}$ is the transmission range adjustment probability of node $i$

and $0 \leq p_{si} \leq 1$.

As shown in [47], the probability that the node number is n in the coverage area of node is Poisson distribution; therefore, the probability that there are $n$ nodes in the survival area is:

$$P(n) = \frac{(\rho\Delta)^n}{n!} e^{-\rho\Delta} \tag{25}$$

where $\Delta$ is the survival area that shown in Fig. 4; $\rho$ is the node density of the network; $n$ is the RND in the candidate relays set.

According to (25), the probability that the RND is in the region $(n_1, n_2)$ can be calculated as:

$$P(n_1 \leq n \leq n_2) = \sum_{n=n_1}^{n_2} \frac{(\rho\Delta)^n}{n!} e^{-\rho\Delta} \tag{26}$$

When $P'(n_1 \leq n \leq n_2)|_\Delta = 0$, the (26) will get the maximum value, which means the under this assumption the probability that there are n nodes in the survival area $\Delta$ is the highest. Moreover:

$$P'(n_1 \leq n \leq n_2)|_\Delta = \sum_{n=n_1}^{n_2} \frac{(\rho\Delta)^{n-1}}{n!} (n\rho e^{-\rho\Delta} - \rho^2 \Delta e^{-\rho\Delta}) \tag{27}$$

when $P'(n_1 \leq n \leq n_2)|_\Delta = 0$, we can get the optimal survival area $\Delta^*$.

In Fig. 4, the transmission range of node A is r and the distance between node *A* and node *B* is *d*. So the line *BC* is also equal to *d*. Then the triangle *ABC* is an isosceles triangle. Therefore, the angle of $\theta_{BAC}$ can be calculated as:

$$\theta_{BAC} = \arccos\left(\frac{r}{2d}\right) \tag{28}$$

Then the optimal value of the transmission range can be calculated as:

$$\Delta^* = r^2 \arccos\left(\frac{r}{2d}\right) \tag{29}$$

where $r^2 \arccos\left(\frac{r}{2d}\right)$ is the area of the survival area.

Based on (27) and (29), we can get the optimal transmission range $r^*$ for the source node. This transmission range has the highest probability to guarantee that the RND in candidate relays set is in the region $(n_1, n_2)$. Therefore, when the node is in unhealthy region, the node transmission ranges will be adjusted to $r^*$.

Therefore, according to (24) and Definition 2, we can calculate the transmission range adjustment ratio of OTC as follows:

$$C_{OTC} = P_u + \frac{1}{N} \sum_{i=1}^{NP_s} p_{si} \tag{30}$$

where $P_u$ means the probability that the node is in unhealthy region; $p_{si}$ is the transmission range adjustment probability of node *i*; $N$ is the node number in network. According to (25) and (26), the $P_u$ can be calculated as:

$$P_u = P[(0 \leq n \leq n_1) \cup (n_2 \leq n \leq N)] \\ = \sum_{n=0}^{n_1-1} \frac{(\rho\Delta^*)^n}{n!} e^{-\rho\Delta^*} + \sum_{n=n_2+1}^{N} \frac{(\rho\Delta^*)^n}{n!} e^{-\rho\Delta^*} \tag{31}$$

Note that there are two variable application-specific parameters in OTC, which are the boundary values of the healthy region: n1 and n2 (i.e. P1 and P2). Since different applications have different network parameters and QoS (Quality of Services) requirements, so these two parameters are not fixed; they are different with different applications. For instance, if the application requires that the packet delivery ratio between the sender and the candidate relays set should larger than 0.9, then the healthy region can be decided as $(0.9, 0.99)$, since the probability can not equal to 1; then the RND can be calculated based on these two probabilities and (20).

**C. Cross-layer and reliable opportunistic routing algorithm**

Based on the link lifetime predication algorithm and the OTC algorithm, in this section, we propose the cross-layer and reliable opportunistic routing algorithm for MANETs.

In CBRT, each node knows the locations of other nodes in the network. If the source node wants to send data packet to the destination node, first, it will broadcast the RREQ to its neighbors. The neighbors are the nodes who have one-hop bi-directional communication links with the source node. The RREQ includes the locations of the source node and the destination node. The neighbors who receive this message will calculate the distances to the destination node. The neighbor nodes whose distances to the destination node are larger than the distance between source node and destination node will drop the RREQ packet directly. Only the neighbor nodes whose distances to the destination node are smaller than the source node will reply RREP message to the source node. In RREP, the concerned cross-layer parameters (including the moving speed and moving direction) are included.

When the source node receives the RREP packet, it will calculate the RND in the candidate relays set and decide which region that it belongs to according to the value of RND. The source node adjusts its transmission range by using OTC algorithm. After the transmission range adjustment, the source node will update the candidate relays set. If the optimal transmission range is smaller than the previous one, then the source node deletes the nodes whose distances to the source node are larger than the transmission range from the candidate relays set. If the transmission range is larger than the previous one, the source node needs to broadcast RREQ packet again. Only the nodes which are not in the candidate relays set and the distances to the destination node are smaller than the source node reply RREP message to source node. After the candidate relays set updating, the source node extracts the cross-layer parameters from the RREP packets. According to (6), the source node calculates the relative variance $D_i$ of each cross-layer parameters. These relative variances $D_i$ are the inputs of the fuzzy logic inference system. The source node relays the data packet to the nodes in the candidate relays set. The transmission priorities of the relaying nodes in the candidate relays set are decided by the utilities that calculated by (12). When the relaying nodes receive the data packet, they will repeat the transmission process that introduced above till the data packet is received by the destination node.

The process of the cross-layer and reliable opportunistic routing algorithm is shown as follows:

**Algorithm 1**. Cross-layer and reliable opportunistic routing algorithm
1. source node broads RREQ to its neighbor nodes;
2. **if** distance(node_i, destination_node) < distance(source_node, destination_node);
3. node *i* send RREP to source node;
4. source node calculates the RND by using OTC algorithm;
5. **end if**
6. **if** ($U_i^j \geq U_i^{th}$)
7. node_status ← true;
8. **else if** ($U_i^j < U_i^{th}$)

9. node_status ← false;
10. **end if**
11. **while** node_status == true do
12. **if** $0 < RND < n_1$ or $n_2 < RND < N$
13.     $p_{si} = P_{si}(RND)$;
14.     $r_i \xleftarrow{p_{si}} r^*$;
15. **else if** $n_1 < RND < n_2$
16.     $p_{si} = 0$;
17.     $r_i = r_i$;
18. **end if**
19. source node update the candidate relays set;
20. source node extract cross-layer parameters from RREP;
21. $D_i \leftarrow rv(U_i^j)$;
22. $\omega_i \leftarrow$ Fuzzylogic($D_i$);
23. $\psi \leftarrow$ Utility($R^1_{Metric\_j}, \omega_i$);
24. ranking($\psi$);
25. end while
26. source node sends data packet to the relaying nodes with the priority list;
27. relaying nodes relay the data packet according the priority in the priority list;
28. The relaying nodes repeat 1-28 till the data packet received by the destination node.

## V. SIMULATION AND DISCUSSION

In this section, the CBRT algorithm will be evaluated. The performance of CBRT algorithm will be compared with the ExOR algorithm in this section. The effectiveness of SBFL algorithm has been demonstrated in Section III; so in the first part of this section, the performance of the packet delivery ratio based topology control algorithm is presented; in the second part of this section, the performance of CBRT and ExOR will be compared in detail. The simulation tool is NS-2.

### A. Performance of packet delivery ratio based opportunistic topology control algorithm (OTC)

In this section, the performance of the packet delivery ratio based opportunistic topology control algorithm will be evaluated.

In this simulation, we compare the OTC with the traditional k-connection algorithm which is the popular topology control algorithm in wireless sensor and ad hoc networks [48][49][50]. In this simulation, the boundary values of the different regions are the same as that show in Section IV; the value of k in k-connection algorithm is equal to 5. The simulation results can be found in Fig. 8 and Fig. 9. The parameters of the simulation are: deployment area: 1000m×1000m; initial node transmission range: 100m; packet length: 1024bits; data rate: 15Kbps; initial energy: 5J; high transmission power: 0.8W; low transmission power: 0.1W; node number: 50-200; average moving speed: 0.2m/s; receiving power: 0.05W.

From Fig. 8, we can conclude that the transmission range adjustment ratio of OTC is smaller than that of k-connection algorithm. With the increasing of the node number, both the adjustment ratios of OTC and k-connection increase. Additionally, when the number of node increases, the difference of the adjustment ratios between OTC algorithm and k-connection algorithm decreases. In Fig. 9, the node degrees of k-connection algorithm vary around 5, which is consistent with the value that we set in this simulation. In OTC algorithm, the node degree varies from 7 to 9 which are the boundary values that calculated in Section IV. The results in Fig. 9 illustrates that the OTC algorithm is effective on controlling the network topology.

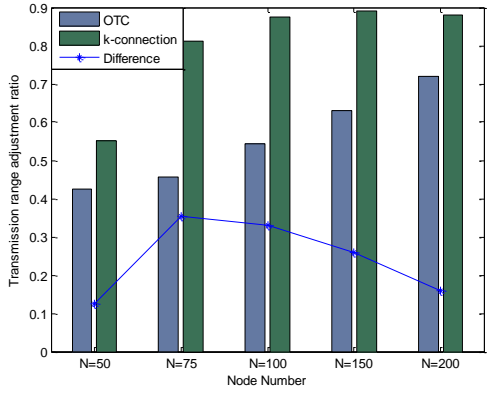

Fig. 8. The transmission range adjustment ration of OTC and *k-connection* algorithm.

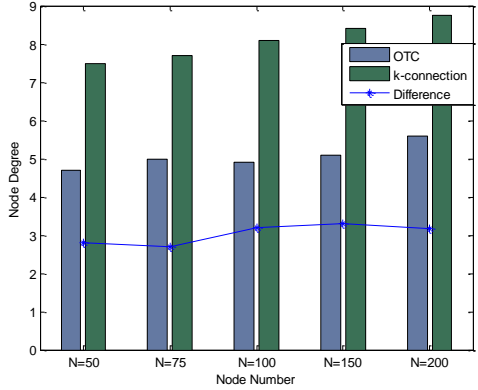

Fig. 9. The node degree of OTC and *k-connection* algorithm.

**B. Performance of the cross-layer and reliable opportunistic routing algorithm (CBRT)**

As demonstrated in Section III, since the SBFL algorithm can handle more cross-layer parameters than the traditional algorithms without increasing the computation complexity, so in this simulation, the residual energy of the relaying node, the ETX of the communication link, the packet queue length in the relaying node, the delay of the relaying node, the distance to the destination node, the moving speed, and the moving direction will be taken into account to choose and prioritize the relaying nodes for the opportunistic routing. It should be noted that the numbers and kinds of the parameters used in this algorithm are not fixed; they can be changed according to the different applications easily and conveniently. This is also the advantage of CBRT algorithm, which is flexible without increase the algorithm complexity.

In this section, we compare the performance of CBRT with the traditional opportunistic routing ExOR [20]. The results can be found from Fig. 10 to Fig. 17. The parameters of the simulation are: deployment area: 1000m×1000m; initial node transmission range: 500m; packet length: 1024bits; data rate: 15Kbps; initial energy: 5J; high transmission power: 0.8W; low transmission power: 0.1W; node number: 25-150; average moving speed: 0.2m/s; receiving power: 0.05W.

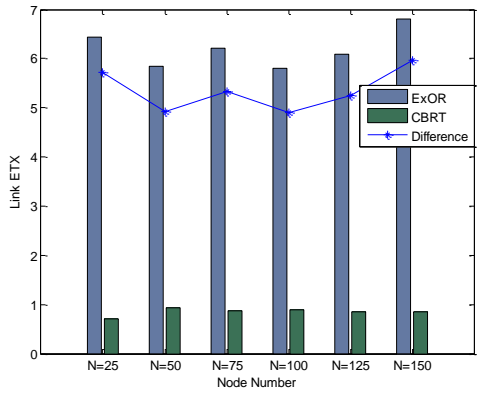

Fig. 10. The ETX of ExOR and CBRT.

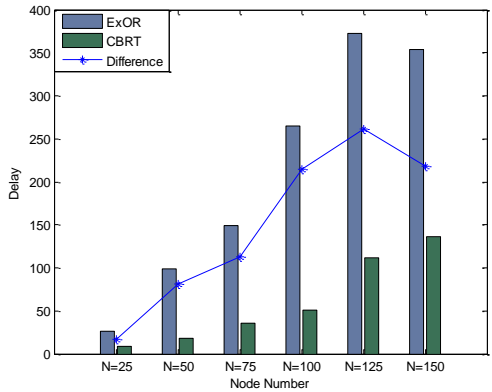

Fig. 11. The transmission delay of ExOR and CBRT.

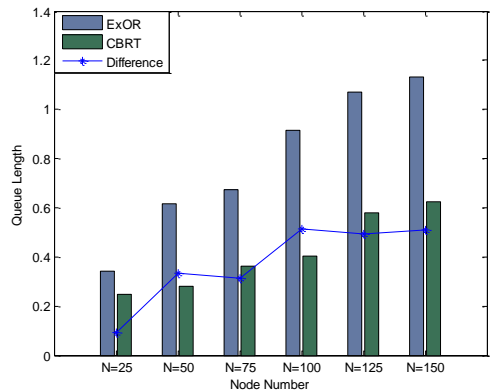

Fig. 12. The queue length of ExOR and CBRT.

In Fig. 10, the ETX of CBRT and ExOR are compared. From Fig. 10, we can find that with the increasing of the node number, the ETX in ExOR varies between 6 and 7; however, this value is only 1 in CBRT. This conclusion demonstrates that in ExOR, the ETX is about 6 times larger than that in CBRT. In MANETs, small ETX means small transmission delay, which can be found in Fig.11. In Fig. 11, with the increasing of the node number, both the transmission delays in CBRT and ExOR increase. However, the increasing in ExOR is more serious than that in CBRT. This is due to the large ETX of ExOR. Not only the large ETX but also the packet queue length in the node can increase the transmission delay. As shown in Fig. 12, the packet queue length in ExOR is also larger than that in CBRT. With the increasing of the node number, the queue length increases both in ExOR and CBRT algorithm. However, the queue length increases faster in ExOR than that in CBRT. The Fig. 11 and Fig. 12 can be explained by each other: the large packet queue length means large transmission delay; conversely, the large transmission delay makes the increasing of the queue length. Moreover, as shown in Fig. 10 and Fig. 12, large ETX and packet queue length deteriorate the performance of transmission delay in ExOR.

The Fig. 10, Fig. 11 and Fig. 12 demonstrate that the network performance has been improved greatly by using CBRT algorithm. As shown in Fig. 10, Fig. 11, and Fig. 12, the parameters in the network are not alone; they relate to and can affect each other. Therefore, the more cross-layer parameters are taken into account, the better network performance is. Since the CBRT have the ability to handle multi-parameters without increasing the complexity of the algorithm, so the CBRT is more efficient than the traditional opportunistic routing algorithm.

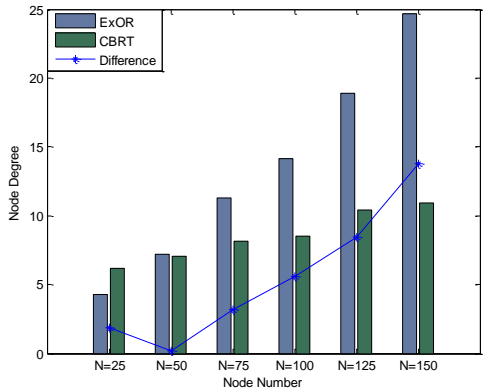

Fig. 13. The node degree of ExOR and CBRT.

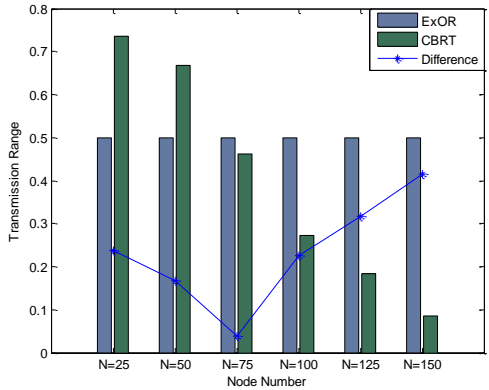

Fig. 14. The transmission range of ExOR and CBRT.

As shown in Fig. 13, due to the topology control in CBRT, the RND in CBRT varies between 6 and 10, which is in the healthy region that we define. In Fig. 13, when the node number increases, due to lack of the topology control in ExOR, the node degree varies greatly in ExOR. For instance, when the node number is 150, the node degree in ExOR is about 5 times larger than that when the node number is 25. The reason is that the transmission range is constant in ExOR, so when the node number increases, the node degree will increase greatly. This also means that the packet delivery ratio varies greatly in ExOR. However, due to the topology control in CBRT, when the node number increases, for maintaining stable RND in the candidate relays set, the transmission range will be adjusted according to the value of RND, which can be found in Fig. 14. In Fig. 14, when the node number increases, due to the OTC algorithm, the transmission range of CBRT decreases to maintain stable RND in the candidate relays set, i.e. the more nodes in the network, the smaller transmission range is. So in Fig. 13, the RND in CBRT algorithm varies slightly, which is in the region that we set in Section IV.

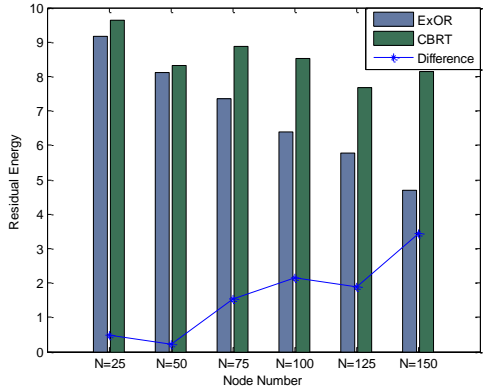

Fig. 15. The residual energy of ExOR and CBRT.

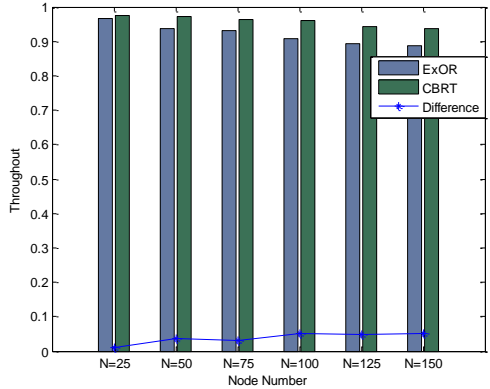

Fig. 16. The through of ExOR and CBRT.

Another advantage of CBRT is the energy consumption. As shown in Fig. 15, the residual energy in CBRT is larger than that in ExOR. With the increasing of the node number, the residual energy decreases both in ExOR and CBRT. However, the energy is consumed faster in ExOR than that in CBRT. The excellent performance of energy consumption in CBRT owes to the balanced performance of the cross-layer parameters, which can be found in Fig. 10, Fig. 11, Fig. 12, and Fig. 14. The performance of ETX, transmission delay, packet queue length, and transmission range in CBRT deduces the energy consumption greatly. Similar to the residual energy, in Fig. 16, with the increasing of the node number, the throughout both in ExOR and CBRT decreases; moreover, the decreasing in ExOR is faster than that in CBRT. This is also because the balanced performance of cross-layer parameters in CBRT.

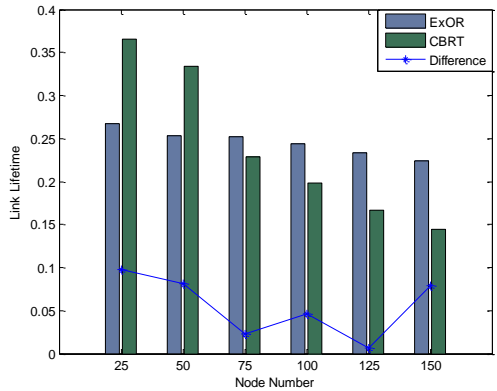

Fig. 17. The residual link lifetime of ExOR and CBRT.

As shown in Fig. 17, the only disadvantage of CBRT is that with the increasing of the node number, the link lifetime in CBRT decreases; however, the link lifetime varies slightly in ExOR. This can be explained by Fig. 14. As shown in Section IV, when the moving speed is stable, the larger transmission range, the smaller link lifetime is. Moreover, in CBRT, when the node number increases, the transmission range decreases, which means small link lifetime. However, since the transmission range is not adjusted in ExOR, so the link lifetime keeps steadily.

## VI. CONCLUSION

In this paper, we propose cross-layer and reliable opportunistic routing algorithm (CBRT) for MANETs. In CBRT, the RND in the candidate relays set is a range rather than a constant number. The node is divided into three regions based on the value of RND. The nodes adjust their transmission range according to the RND in the candidate relays set. The cross-layer metrics are not inputted into the fuzzy logic system directly; the inputs are the relative variances of the metrics. The calculation of the node utility is weight based. The weight and the parameter sequence in the metric are used to calculate the node utility. The relaying node which the node utility is high will have high priority to relay the data packet. By these innovations, in CBRT, the network performance is much better than that in ExOR; however, the computation complexity is not increased.